\documentclass[sigconf,nonacm]{acmart}
\renewcommand\footnotetextcopyrightpermission[1]{}
\usepackage{graphicx}
\usepackage{amsmath}
\usepackage{mathtools}
\usepackage{xspace}
\usepackage{array}
\usepackage{booktabs}
\usepackage{multirow}
\usepackage{xcolor}
\usepackage{textcomp}
\usepackage{algorithm}
\usepackage{algorithmic}

\newcommand{\sw}{\textsc{ScratchWorld}\xspace}

\long\def\com#1{}

\makeatletter\def\@parfont{\bfseries\itshape}\makeatother
\newcommand{\para}[1]{\paragraph{#1}}
\definecolor{swBlue}{HTML}{2a78d6}
\definecolor{swOrange}{HTML}{eb6834}
\definecolor{swGreen}{HTML}{008300}
\definecolor{swInkTwo}{HTML}{52514e}
\definecolor{swViolet}{HTML}{4a3aa7}
\definecolor{swMagenta}{HTML}{e87ba4}
\definecolor{swYellow}{HTML}{eda100}
\definecolor{swAqua}{HTML}{0e9c6c}
\definecolor{swRose}{HTML}{c94f7c}
\definecolor{swGray}{HTML}{6b6a66}
\newcommand{\hlgate}[1]{\colorbox{swOrange!14}{#1}}
\newcommand{\hlva}[1]{\colorbox{swBlue!13}{#1}}
\newcommand{\hlladder}[1]{\colorbox{swGreen!11}{#1}}
\newcommand{\algcmt}[1]{{\unskip\nobreak\hfil\penalty50\hskip1em\hbox{}\nobreak\hfil
  \mbox{\scriptsize\color{swInkTwo}$\triangleright$ #1}%
  \parfillskip=0pt \finalhyphendemerits=0 \par}}

\usepackage{tcolorbox}
\tcbuselibrary{skins,breakable}
\newtcolorbox{findingbox}{
  enhanced, breakable,
  colback=black!6, colframe=black!6,
  boxrule=0pt, arc=2pt,
  left=5pt, right=5pt, top=4.5pt, bottom=4.5pt, boxsep=0pt,
  before skip=11.5pt plus 2pt, after skip=11.5pt plus 2pt,
}
\newcommand{\finding}[2]{%
\begin{findingbox}%
\textbf{Finding (#1):} #2%
\end{findingbox}}

\AtBeginDocument{\setlength{\emergencystretch}{3em}}
\AtBeginDocument{\brokenpenalty=100}
\AtBeginDocument{\setlength{\parskip}{0pt plus 1pt}}

\begin{document}

\title[ScratchWorld: Evaluating If World Models Compute Executable Consequences]{\sw: Evaluating If World Models\texorpdfstring{\\}{ }Compute Executable Consequences}

\author{Yufeng Lin}
\affiliation{%
  \institution{Pui Ching Middle School Macau}
  \city{Macau}
  \country{China}}

\author{Jialu Zhang}
\authornote{Corresponding author.}
\affiliation{%
  \institution{University of Waterloo}
  \country{Canada}}

\renewcommand{\shortauthors}{Lin and Zhang}

\begin{abstract}
Evaluations of world models commonly score a predicted future state by its overlap with the ground truth. 
Yet in interactive worlds, this rewards copying: a model that merely preserves persistent state appears accurate without predicting any consequence of an action. 
We introduce \sw, an execution-grounded diagnostic benchmark that treats Scratch projects as executable worlds and derives every label from replay on a pinned Scratch VM, including next-state transitions, hidden variables, causal traces, and counterfactual outcomes. Its primary metric, value-aware changed-field $F_1$ ($F_1^{\mathrm{VA}}$), credits a prediction only when it names a changed field and computes its executed value. 
Model-free anchors make the confound explicit: a no-change prior attains $91.4\%$ full-state field accuracy on synthetic long-horizon items and $0\%$ changed-field credit on non-empty one-step items, whereas VM replay attains $100\%$ wherever the corresponding metric is defined. 
Across seven frontier LLMs prompted as world models, state-only $F_1^{\mathrm{VA}}$ reaches at most $30.0\%$ on controlled synthetic mechanisms and declines steeply as projects approach real-world complexity. Revealing the full program raises every model's value-aware score by $22.3$--$51.7$ points and substantially reorders the models, yet the best reaches only $55.4\%$, an execution gap that persists even with the rules visible. Auxiliary diagnostics complete the error profile: causal-event accuracy remains within $2.5$ points of the $50\%$ no-event baseline, with no correct exact chain; counterfactual prediction trades changed-outcome accuracy against invariance; hidden-state accuracy degrades roughly twice as fast as full-state overlap over long horizons. In a word, \sw separates what a model knows about the rules from what it can actually execute.
\end{abstract}


\maketitle

\hypersetup{
  pdftitle={ScratchWorld: Evaluating If World Models Compute Executable Consequences},
  pdfauthor={Yufeng Lin, Jialu Zhang},
  pdfkeywords={world models, executable worlds, Scratch, execution-grounded evaluation, causal reasoning, counterfactual reasoning}
}

\section{Introduction}
\label{sec:intro}

A world model is useful only if it predicts how the environment changes when an agent acts~\citep{ha2018world,hafner2023dreamerv3}. The standard way to score such predictions is overlap: compare the predicted future state with the ground truth and count what matches. Yet in interactive environments, most state variables remain unchanged after a single action; a single action typically alters a few fields while hundreds persist, so a model that simply copies the present already matches almost everything. Overlap therefore rewards persistence at least as much as prediction. We study this confound in program-defined worlds whose dynamics are executable, inspectable, and replayable, so that the confound can be controlled exactly. What we find is a severe capability gap. Even with the governing program in full view, frontier models largely fail to execute the consequences of an action on the world's state.

\begin{figure}[t]
\centering
\includegraphics[width=\linewidth]{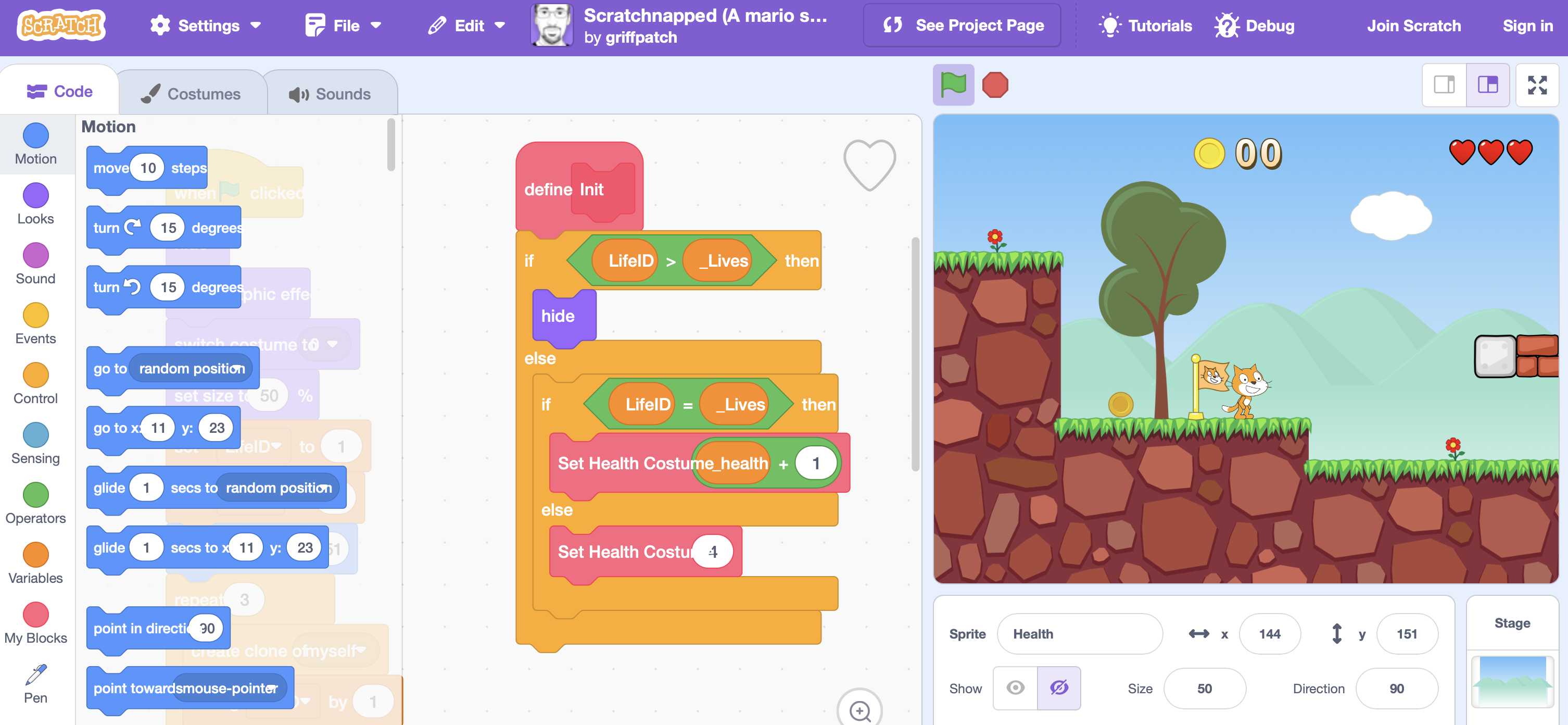}
\caption{An example Scratch project simulating a Mario game. Source: \url{https://scratch.mit.edu/projects/10118230/}.}
\label{fig:scratch-gui}
\end{figure}

Existing evaluations only partially isolate this capability. Video benchmarks provide rich observations, but actions, hidden state, and inexpensive counterfactuals are often unavailable or weakly specified~\citep{yi2020clevrer,huang2024vbench}. Robotics and embodied-agent benchmarks provide action-conditioned trajectories, but data collection and controlled interventions are expensive~\citep{dasari2019robonet,savva2019habitat}. Game benchmarks expose interaction, but score reward, task success, or observation quality rather than the transition itself~\citep{cobbe2020procgen}. All three settings inherit the persistence baseline familiar from forecasting and video prediction: when state changes slowly, repeating the last observation is strong~\citep{rasp2020weatherbench,mathieu2016deep}. A high overlap score may therefore reward persistence rather than prediction, while hiding whether a model failed to predict changes, hidden variables, causal events, or intervention outcomes.

We introduce \sw, an execution-grounded benchmark that turns Scratch projects into action-conditioned world-model tests. Scratch offers a combination that few existing testbeds provide: an ecosystem of user-authored projects with real behavioral diversity; fully inspectable hidden state (variables, lists, clones, pending broadcasts, active scripts) behind every rendered frame; a deterministic Virtual Machine (VM) whose few sources of nondeterminism can be pinned, so that any transition, causal trace, or counterfactual intervention replays exactly in milliseconds; and a renderer that exposes only a lossy view of that state, so observation alone cannot substitute for state tracking. From a curated corpus of real and synthetic projects, \sw compiles $892$ replay-verified instances across four tasks, one per failure mode above: next-state prediction, causal event attribution, counterfactual prediction, and long-horizon state tracking. Each label is an executed action--state--outcome fact, not a human annotation.

Scoring these predictions must resist the confound directly. A Scratch state contains many fields and a local action changes few, so \sw adopts value-aware changed-field $F_1$ ($F_1^{\mathrm{VA}}$) as its primary metric: credit requires both naming a changed field and computing its executed value. A model-free evidence ladder, comprising a copy prior, an action heuristic, and a VM oracle, quantifies exactly how much score persistence alone can buy: on different task-specific pools, the copy prior attains $91.4\%$ full-state field accuracy on synthetic long-horizon items and $0\%$ changed-field credit on non-empty one-step items, the action heuristic peaks at $10.3\%$ even on controlled synthetic mechanisms, and the VM oracle attains $100\%$ wherever the corresponding metric is defined. Presence-only $F_1$ ($F_1^{\mathrm{pres}}$) and no-op accuracy isolate field selection and invariant outcomes.

Across seven frontier models, three findings define the picture. First, \emph{real projects defeat every model}: value-aware $F_1$ on real-project changed outcomes is $0.0\%$ for all seven. The strongest systems (\textsc{Opus-5}, \textsc{GPT-5.6-Sol}) occasionally name a correct changed field but never compute its executed value, even though the same models reach up to $30.0\%$ on synthetic worlds. Second, \emph{missing rules and failed execution are separate bottlenecks}: revealing the full program raises every model by $22.3$--$51.7$ points, compresses the cross-model range from $30.0$ to $13.4$ points using unrounded scores, and reorders the ranking. Yet the best model still reaches only $55.4\%$ against the $100\%$ VM oracle, and its residual errors include omitted and invented fields, not just wrong values. Third, \emph{causal grounding is absent}: on a balanced slice, event accuracy remains within $2.5$ points of the $50\%$ no-event baseline, and no model recovers a single correct causal chain. Auxiliary probes further sharpen the error profile: hidden-state accuracy deteriorates roughly twice as fast as full-state overlap over twenty actions; counterfactual evaluation reveals a trade-off between sensitivity and invariance across models; and reasoning effort and tool use produce model-dependent effects rather than consistent gains.
 
These results argue that executable consequence prediction should be reported as an error profile rather than a single overlap number, along five separable axes: field selection, value execution, rule access, event grounding, and hidden-state tracking. Because every axis is replay-verified, the same pipeline that diagnoses these failures can provide a foundation for developing more reliable training and evaluation methods for world models.

\para{Contributions.}
We make four contributions:
\begin{itemize}
\item We introduce \sw, a replay-grounded pipeline turning Scratch programs into action-conditioned next-state, causal, counterfactual, and long-horizon instances with hidden state.
\item We define value-aware changed-field evaluation and a model-free evidence ladder (copy prior, action heuristic, VM oracle) separating persistent-state overlap from executed change.
\item We use matched state-only and state$+$program evidence to distinguish missing-rule errors from residual execution errors, showing large gains, a compressed score range, substantial rank reordering, and a remaining $44.6$-point gap to exact VM execution.
\item We conduct a systematic diagnostic study of seven frontier models, revealing a consistent execution gap across synthetic and real worlds, including failures in value execution, causal grounding, counterfactual reasoning, and long-horizon hidden-state tracking.
\end{itemize}
\section{Background}
\label{sec:background}

\paragraph{Scratch interface and semantics.}
Scratch is a block-based visual programming environment in which users compose scripts by stacking connected blocks~\citep{resnick2009scratch}; Figure~\ref{fig:scratch-gui} shows the editor and the stage on which a project runs. A project is serialized into a structured representation together with its media assets and executed by the Scratch Virtual Machine (VM). Events such as the green flag, user inputs, and broadcasts trigger scripts, and the VM schedules concurrent scripts under a well-defined event-driven execution model, making program behavior reproducible. Each sprite carries both visible attributes (position, costume, visibility) and hidden runtime state (variables, lists, clones, pending broadcasts, active scripts); clones inherit their parent's scripts and state at creation, so a single script can govern many dynamically created entities. Apart from explicit sources of nondeterminism (random-number blocks, clock reads, and runtime-assigned identifiers), execution is fully determined by the program and its inputs, yet its state is only partially exposed by the rendered frame.

\paragraph{Projects as action-conditioned worlds.}
This design maps directly onto the vocabulary of world models. The serialized runtime snapshot is a state; user inputs arriving in a window are actions; VM ticks define the transition function; and the rendered frame is a lossy observation of the underlying state. Because the few nondeterministic inputs are explicit, they can be pinned, and any transition, causal trace, or counterfactual outcome can then be reproduced exactly by rerunning the VM. Interventions are equally inexpensive: editing a variable, a threshold, or an event condition and replaying costs milliseconds, whereas the corresponding controlled experiment in real-world or high-fidelity simulation environments is expensive or even impossible. The combination of visible behavior, inspectable program structure, and replayable runtime state makes Scratch a suitable substrate for evaluating whether models can compute action-conditioned consequences.
\section{Motivating Example}
\label{sec:motivating-example}

\begin{figure*}[t]
\centering
\includegraphics[width=\textwidth, trim=0 0 0 3, clip]{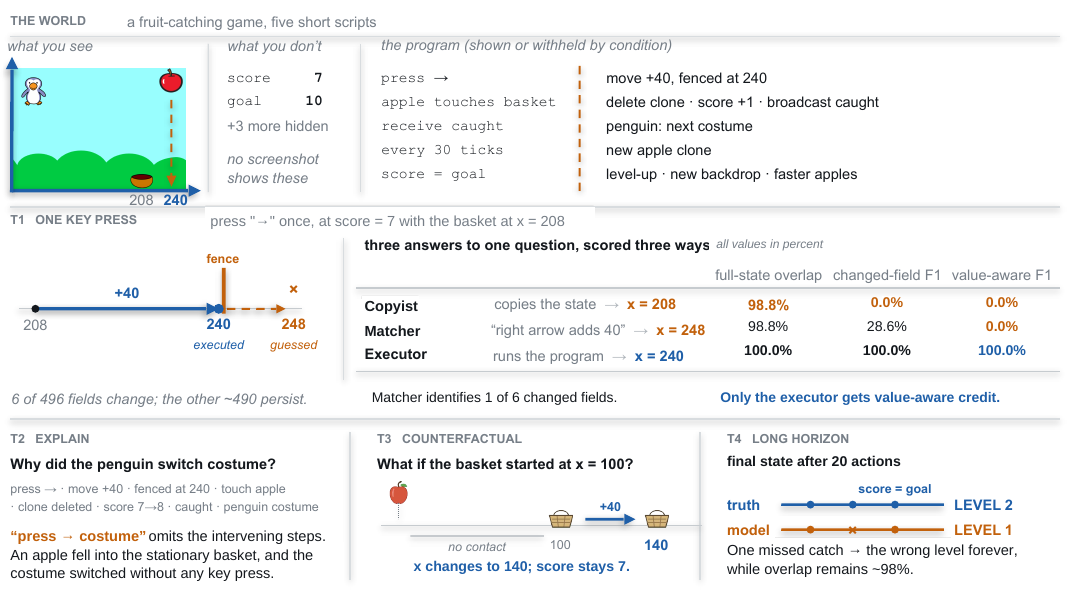}
  \caption{One press, three answers, three verdicts.
  \textit{Top}: the stage, the hidden state that no screenshot reveals, and the five
  scripts that generate every transition; an evidence condition controls how
  much of the program a model sees.
  \textit{Middle} (T1): the fence clamps the basket at $x{=}240$, not $248$.
  Full-state overlap makes the do-nothing copyist look nearly perfect ($98.8\%$);
  presence-only changed-field $F_1$ credits the matcher for identifying only one
  of the six changed fields, while value-aware changed-field $F_1$
  ($F_1^{\mathrm{VA}}$) gives credit only to the executor.
  \textit{Bottom}: the executed chain explains the costume switch (T2),
  an intervention removes the collision-mediated consequences (T3), and one
  missed catch flips the regime forever while overlap stays near $98\%$ (T4).}
\label{fig:world}
\end{figure*}

Consider a fruit-catching game of five short scripts
(Figure~\ref{fig:world}): pressing the right arrow moves the basket $40$
steps, and the runtime \emph{fences} sprites at the stage edge; an apple
touching the basket is deleted, \texttt{score} rises by $1$, and
\texttt{caught} is broadcast; a penguin that receives \texttt{caught}
switches costume; when \texttt{score} reaches \texttt{goal}${=}10$,
\texttt{level-up} switches the backdrop and speeds up the apples; every
thirty ticks a new apple spawns. The rules can be understood within a minute. The rendered frame shows two sprites and a falling apple;
\texttt{score}, \texttt{goal}, the clone count, and the pending
broadcasts are hidden. Because the dynamics are a program, every
question below has exactly one executable answer. How much of that
program a model may inspect is an \emph{evidence condition}; the
experiments vary it from hidden to fully visible.

Press the right arrow once, at \texttt{score}${=}7$ with the basket at
$x{=}208$. The move block adds $40$, but the fence stops the basket at
$x{=}240$, not $248$; it now touches the apple, so the clone is
deleted, \texttt{score} becomes $8$, \texttt{caught} fires, and the
penguin switches costume. Six fields change; the remaining $490$ of this
$496$-field state persist. 
A \emph{copy-prior predictor} that repeats the input state achieves 98.8\% full-state overlap.
A \emph{pattern matcher} that has learned ``right arrow adds $40$''
answers $x{=}248$: the right field, an unexecuted value. It also gets $98.8\%$ full-state overlap and $28.6\%$ presence-only $F_1$, but $0\%$ value-aware $F_1$. Only an executor produces $x{=}240$ and \texttt{score}${=}8$. Full-state overlap therefore makes both non-executors look nearly perfect, while value-aware changed-field $F_1$ gives credit only to the executor. Task~1 therefore scores one-step prediction by the changed fields \emph{and} their
executed values.

Why did the penguin change costume? The executed answer is the chain
above: press, move, collision, deletion, increment, broadcast,
receiver. The tempting answer, ``pressing the arrow changes the
costume,'' is a shortcut learned from one example trajectory, and this world
breaks it both ways: press with the basket at $x{=}100$ and nothing
switches; drop an apple into a stationary basket and the costume
switches without a press. Observing the event is not enough; the mechanism is the test. Task~2 therefore scores the predicted rule chain against
the executed trace.

What if the basket \emph{had} been at $x{=}100$? Replay the press. The
basket lands at $140$ and touches nothing: no deletion, no broadcast,
\texttt{score} stays $7$. The basket still moves from $100$ to $140$, so this is not a full-state no-op;
what disappears are the collision-mediated consequences: no deletion, no broadcast,
\texttt{score} stays $7$, and the costume does not switch. The correlational shortcut above must
hallucinate a costume switch here, and inventing consequences is as
wrong as missing them. A planner lives on exactly such untaken actions.
Task~3 therefore scores the outcome of an intervention, including correctly predicting when it produces no collision-mediated consequence.

Predict the final state after twenty more actions, with no intermediate ground truth. At the third catch \texttt{score} reaches
\texttt{goal}, \texttt{level-up} fires, and the world runs under new
dynamics; a model must switch regimes at exactly that step. Drop one
catch and \texttt{score} is off by one forever: the model simulates
the wrong level from then on, while the hundreds of fields it copies
keep overlap comfortably high. One lost hidden variable falsifies the
entire future, and full-state overlap remains high. Task~4 therefore tracks
state over long horizons, hidden fields included.

One rule chain, four directions: compute a step, explain it, recompute
it under an intervention, sustain it through time. The rest of the
paper is this example at scale, and the question it asks is not
whether a model can produce a plausible next frame, but whether its
predictions follow the executable rules that generate the world.
\section{Related Work}

\paragraph{World models and action-conditioned dynamics.}
World models learn predictive representations of environment dynamics for imagination, planning, and control. Classic and recent systems support behavior through learned dynamics, from latent-dynamics control \citep{ha2018world,hafner2019planet,hafner2020dream,hafner2023dreamerv3} and search with learned models \citep{schrittwieser2020muzero} to scalable continuous-control world models \citep{hansen2024tdmpc2}. Transformer- and diffusion-based agents extend this line in visual environments \citep{micheli2023iris,alonso2024diamond}, while Genie learns controllable generative environments from video \citep{bruce2024genie}. Standard evaluations emphasize reward, task success, or observation prediction. The copy confound we quantify is the executable-world form of the persistence baseline: in weather forecasting and video prediction, repeating the last observation is a well-known strong baseline under slowly changing state, and benchmarks report it explicitly \citep{rasp2020weatherbench,mathieu2016deep}. \sw ports this audit to program-defined worlds, pairs it with a changed-field metric that a persistence predictor cannot satisfy, and focuses on the transition structure behind observed outcomes: hidden variables, executable rules, and counterfactual dependencies.

\paragraph{Physical, causal, and video-generation benchmarks.}
Many benchmarks probe physical and causal reasoning. IntPhys, PHYRE, and Physion evaluate intuitive physics, physical problem solving, and future physical prediction \citep{riochet2018intphys,bakhtin2019phyre,bear2021physion}; CausalWorld provides intervention-controlled robotic tasks for causal structure and transfer \citep{ahmed2021causalworld}; and CLEVRER separates descriptive, explanatory, predictive, and counterfactual reasoning over collision videos \citep{yi2020clevrer}. For generative video, VBench and VideoPhy move from perceptual quality toward physical consistency \citep{huang2024vbench,bansal2024videophy}, while Vista studies high-fidelity, action-controllable future generation for autonomous driving \citep{gao2024vista}. Recent world-model evaluations make related points: World-in-World emphasizes closed-loop controllability \citep{zhang2025worldinworld}, and VisPhyWorld converts predictions into executable simulator code for physics-based checking \citep{liang2026visphyworld}. \sw supplies the executable world directly and tests action-conditioned state prediction, event attribution, and controlled interventions.

\paragraph{Interactive environments, games, and embodied-agent benchmarks.}
Interactive benchmarks couple actions with consequences. Procgen and Crafter evaluate generalization and broad agent capabilities in procedurally generated games \citep{cobbe2020procgen,hafner2022crafter}; Habitat and ALFRED evaluate embodied interaction and instruction following \citep{savva2019habitat,shridhar2020alfred}; RoboNet aggregates large-scale robot interaction data \citep{dasari2019robonet}. These settings expose action-conditioned feedback through reward, success rate, or task completion. Executable text-game benchmarks are closest to our task structure: TextWorld supplies programmatic text environments \citep{cote2018textworld}, Jericho provides human-authored interactive-fiction games \citep{hausknecht2020jericho}, ScienceWorld grounds scientific tasks in an interactive text environment \citep{wang2022scienceworld}, and \citet{wang2024simulators} score LLM state-transition prediction against game-engine ground truth. \sw differs in substrate and supervision: its worlds are visual, event-driven, and concurrently scheduled programs with clones and broadcasts; its labels include VM-derived hidden state, executed causal chains, and replay-verified counterfactuals rather than next-state tuples alone; and its value-aware changed-field scoring separates field identification from value computation, which a whole-state comparison merges. \sw also keeps the transition mechanism inspectable: actions, broadcasts, variables, event handlers, clone states, and field-level updates are logged and checked against the executed program.

\paragraph{Scratch, block-based programming, and executable media.}
Scratch is a block-based, sprite-based, event-driven programming environment \citep{resnick2009scratch}. Prior Scratch work uses the same environment to study different questions. Static-analysis tools such as Hairball and LitterBox detect code patterns and potential bugs \citep{boe2013hairball,fraser2021litterbox}, while testing tools such as Whisker and model-based Scratch testing execute projects to check program behavior against abstract state machines \citep{stahlbauer2019testing,gotz2022modelbased}. ScratchEval studies multimodal visual-programming reasoning with Scratch-style questions \citep{fu2025scratcheval}, and another line evaluates multimodal GUI agents that construct Scratch programs through drag-and-drop \citep{zhang2026seeplansnap}. More generally, CRUXEval evaluates program input/output execution reasoning, and NExT uses runtime traces to train execution-aware reasoning \citep{gu2024cruxeval,ni2024next}. These are code-understanding, analysis, testing, and agent benchmarks: their unit of evaluation is a program to be read, checked, or built. \sw uses a different prediction unit: an existing executable world, a current state, an action window, and a VM-derived target. To our knowledge, it is the first Scratch benchmark to supply VM-verified next-state, hidden-state, causal-chain, and counterfactual targets for the same executable world under a copy-sensitive metric. It treats Scratch as a world-model substrate with program-derived ground truth rather than as an object of program comprehension.

\section{Benchmark Construction}
\label{sec:benchmark-construction}

\begin{algorithm}[t]
\caption{How \sw{} builds and scores one instance. Phase~1 is the pipeline
of this section. Phase~2 is the changed-field scoring rule of the
Evaluation section. Phase~3 provides the model-free anchors of
Table~\ref{tab:metric-diagnostic}. ``$\approx$'' denotes equality within the
numeric tolerance of the Evaluation section; the heuristic anchor ($10.3$)
is its value-aware $F_1$ on the synthetic changed-delta pool.}
\label{alg:scratchworld}
\small
\begin{algorithmic}[1]
\REQUIRE project $P$; pinned VM $V$; seed $\rho$; action schedule
  $\mathcal{A}$; predictor $f$
\item[] {\footnotesize\color{swInkTwo}\textit{Phase 1: replay-verified labels}}
\STATE run $V$ on $(P,\rho,\mathcal{A})$; log trace $\tau=\{(s_t,a_t,o_t,e_t)\}_{t=0}^{T}$
\FORALL{action windows $(s_t,a_t,s_{t+1})$ in $\tau$}
  \STATE $\Delta s^{*} \leftarrow \{(k, v^{*}(k)) : s_{t+1}[k] \neq s_t[k]\}$,
    where $v^{*}(k)=s_{t+1}[k]$ \hfill\algcmt{gold delta}
  \STATE \hlgate{\textbf{keep} only if a second replay reproduces
    $\Delta s^{*}$} \hfill\mbox{\algcmt{\textbf{determinism gate}}}
  \STATE attach evidence $E$: state-only \emph{or} state$+$program
\ENDFOR
\STATE given a target event $q$: $c^{*} \leftarrow$ handler chain executed
  behind $q$ in $\tau$ \hfill\algcmt{causal gold}
\STATE given an intervention $\iota$: $s^{\iota} \leftarrow$ replay of
  $(P,\rho,\mathcal{A})$ under $\iota$ \hfill\algcmt{counterfactual gold}
\item[] {\footnotesize\color{swInkTwo}\textit{Phase 2: value-aware scoring}}
\STATE model predicts a delta $\hat{\Delta s} \leftarrow f(s_t,a_t,E)$
\STATE \hlva{$(k,\hat{v})\in\hat{\Delta s}$ is correct \textbf{iff}
  $k\in\Delta s^{*}$ \emph{and} $\hat{v}\approx v^{*}(k)$}
\STATE $F_1^{\mathrm{VA}} \leftarrow F_1$(correct pairs);\ \
  $F_1^{\mathrm{pres}} \leftarrow F_1$(fields alone)
\STATE report $F_1^{\mathrm{pres}} - F_1^{\mathrm{VA}}$ as the value-computation gap
\item[] {\footnotesize\color{swInkTwo}\textit{Phase 3: model-free anchors}}
\STATE anchors: \hlladder{copy $0$ $\cdot$ heuristic $10.3$ $\cdot$ oracle $100$}
\end{algorithmic}
\end{algorithm}

\begin{figure*}[t]
\centering
\includegraphics[width=\textwidth, trim=8 30 8 80, clip]{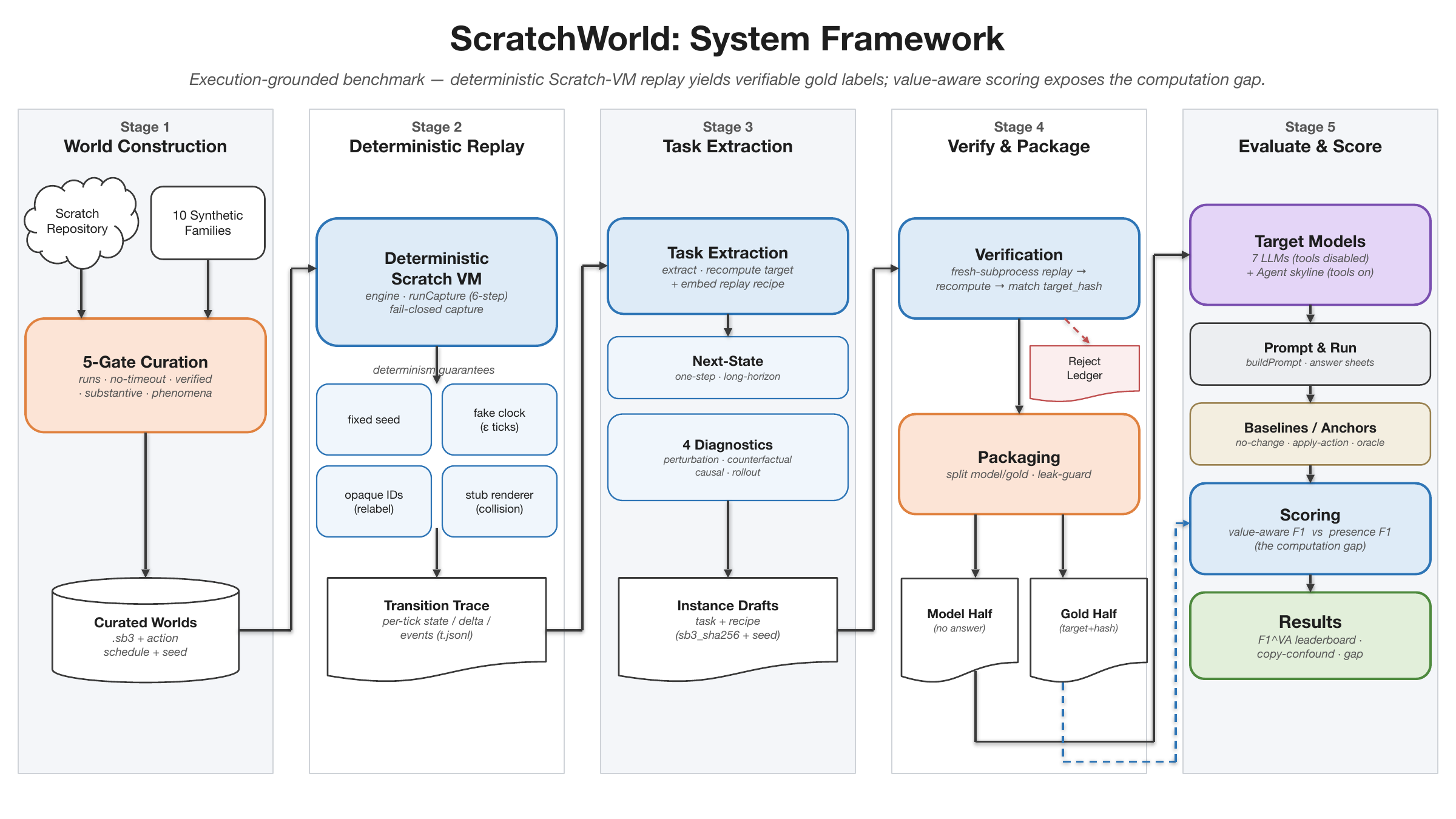}
\caption{\sw{} end-to-end pipeline. \emph{Stage 1 (World Construction)} curates Scratch worlds (a corpus of real projects and ten synthetic mechanism families) through a five-gate filter. \emph{Stage 2 (Deterministic Replay)} runs each world on a pinned Scratch VM under fixed seeds and controlled action schedules. \emph{Stage 3 (Task Extraction)} derives next-state and diagnostic tasks with embedded replay recipes. \emph{Stage 4 (Verify \& Package)} re-verifies every label by fresh-subprocess replay (reject on hash mismatch) and packages a leakage-guarded model/gold split. \emph{Stage 5 (Evaluate \& Score)} runs the prompted models plus model-free anchors and a tool-equipped agent skyline, scoring with value-aware $F_1$ against presence-only $F_1$ to expose the computation gap. Gold labels (dashed blue) are fed only to the scorer, never to the model.}
\label{fig:pipeline}
\end{figure*}

\sw turns Scratch projects into replay-verified executable worlds for evaluating action-conditioned world models (Figure~\ref{fig:pipeline}). The governing principle is that every label is a replay outcome, never a human annotation. Algorithm~\ref{alg:scratchworld} gives the loop end to end: Phase~1 is the construction pipeline described in this section, Phase~2 the scoring rule of the Evaluation section, and Phase~3 the model-free anchors used throughout the Experiments.

We use the following notation for the construction loop. Let $P$ denote a Scratch project, $V$ the pinned VM, $\rho$ the replay seed, $\mathcal{A}$ the injected action schedule, and $f$ the prediction function being evaluated. A replay produces a trace $\tau=\{(s_t,a_t,o_t,e_t)\}_{t=0}^{T}$, where $s_t$ is the structured runtime state, $a_t$ the action or event window applied at step $t$, $o_t$ the rendered observation, and $e_t$ the corresponding runtime-event record; $T$ is the final trace index. We write $k$ for a canonical state field, $E$ for the evidence exposed to the model, $q$ for a target event, and $\iota$ for a controlled intervention. For the replay-derived targets used below, $\Delta s^{*}$ is the gold changed-field set, $v^{*}(k)$ the gold value of field $k$, $c^{*}$ the executed causal chain, and $s^{\iota}$ the counterfactual replay state; hatted quantities denote model predictions.

\para{Projects and filtering.}
We begin with a local corpus of Scratch project files; public projects enter the benchmark only as filtered, derived instances, never as redistributed raw projects. Each project contains sprites, blocks, variables, costumes, sounds, and metadata. We retain projects with meaningful state-changing behavior, including games, simulations, animations, quizzes, drawing programs, and interactive stories, and pass every candidate through the five-gate filter of Figure~\ref{fig:pipeline}: a project is removed if it fails to execute, depends on unsupported VM extensions, contains missing assets, terminates immediately, or behaves unstably under repeated controlled replay. We then remove near-duplicates using structural similarity (block graphs, sprite layouts, event-handler patterns) and surface similarity (titles, names, and visual assets).

\para{Synthetic mechanism families.}
The real corpus is complemented by ten synthetic mechanism families that each isolate a known dynamic: sprite motion, broadcast chains, clone lifecycles, collision rules, delayed effects, hidden variables, list operations, multi-field changes, backdrop and costume switching, and negative controls in which the scheduled action changes nothing. Each family is generated from a parameterized template under seeded randomization, so instances vary in layout, thresholds, and initial state while exercising the same mechanism, and every generated instance passes the same gates and replay verification as the real corpus. The families provide controlled coverage that harvested projects cannot guarantee, and the negative-control family populates the empty-delta pool on which no-op accuracy is evaluated.

\para{Runtime instrumentation and replay.}
For each retained project, we instrument the Scratch runtime to log structured execution traces without changing program behavior. Visible state includes sprite positions, directions, sizes, costumes, visibility flags, and layer order. Hidden state includes global and local variables, lists, clone states, active scripts, broadcasts, and control-flow status. User actions such as green-flag events, key presses, mouse clicks, and sprite clicks are injected through a controlled interface; broadcasts, collisions, threshold branches, and script activations are recorded from the runtime trace. All reported labels use a pinned \texttt{scratch-vm@5.0.300} runtime. 

\para{Determinism and label validity.}
Every retained label is a deterministic function of the project, the pinned VM, the fixed seed, and the injected action schedule: projects that behave unstably under repeated controlled replay are removed during filtering, and each instance must survive an independent replay that reproduces its target exactly (the highlighted gate in Algorithm~\ref{alg:scratchworld}). Every retained instance therefore has a unique, well-defined answer given full program evidence, so the tasks are well-posed at the label level; what an evidence condition withholds from the model is controlled separately (see \emph{Packaging} below and the Evaluation section).

\para{Instance construction.}
From traces we extract one-step transitions $(s_t,a_t,s_{t+1})$, long-horizon transitions $(s_t,a_{t:t+H-1},s_{t+H})$, causal event windows, and counterfactual pairs. Empty-delta outcomes are valid executable labels: for state-delta tasks, the correct prediction is an empty change set when replay shows that no listed field changed. We retain no-change outcomes in counterfactual diagnostics because invariance under a controlled intervention is itself a meaningful target. For causal attribution, ordered runtime traces link a target event to the event handler, condition, broadcast, or script activation that produced the observed change; chain labels are extracted mechanically from the executed block sequence, with no human annotation in the loop. Negative examples are retained when the event does not occur; the correct behavior is then to predict non-occurrence and abstain from inventing a rule chain. Instances whose causal chains cannot be recovered reliably under replay are excluded.

For counterfactual instances, we construct paired executions that differ by one controlled intervention. Interventions may replace an action, modify an initial variable, move a sprite, or change a threshold or event condition. We replay the intervened execution on the same VM and score against the resulting counterfactual state. Each intervened instance receives its own replay-verified target.

\para{Packaging.}
Each instance is packaged under five input modalities: raw project JSON, structured state tables, natural-language summaries, rendered screenshots, and short execution clips. The same replay target can be presented under any modality, so performance can be disaggregated by input modality. This design separates missing information from errors in simulating executable dynamics. The model-facing package never contains targets, gold deltas, or verification metadata; gold labels travel in a separate scorer-only package, so leakage is prevented structurally rather than by prompt discipline.
\section{Evaluation}
\label{sec:eval}

\sw evaluates whether a model can execute the consequences of actions through four tasks: next-state prediction, causal event attribution, counterfactual prediction, and long-horizon state tracking. The tasks move from local transition modeling to causal and temporal reasoning. Throughout, $\mathcal{W}$ denotes the executable project, and the evidence condition specifies which part of that project the model can inspect.

\paragraph{Task 1: Next-state prediction.}
Given a Scratch world $\mathcal{W}$, current state $s_t$, action or event window $a_t$, and optional visual observation $o_t$, the model predicts the state after execution:
\[
\hat{s}_{t+1} = f(\mathcal{W}, s_t, a_t, o_t).
\]
The prediction is compared with the execution-derived ground truth $s_{t+1}$. Outputs take the form of a full structured state or a state delta $\Delta s_t = s_{t+1} - s_t$; an empty delta is correct only when replay shows no listed field changed.

\paragraph{Changed-field scoring.}
The primary metric is \emph{value-aware changed-field} $F_1$ ($F_1^{\mathrm{VA}}$), designed to distinguish predicting change from merely matching persistent state. A prediction receives credit when it names a changed field and predicts its value within tolerance (Algorithm~\ref{alg:scratchworld}, Phase~2). Presence-only changed-field $F_1$ ($F_1^{\mathrm{pres}}$) credits the field name alone. Full-state field accuracy measures overlap with the entire state and serves as a diagnostic for the persistence bias. Field identities are canonicalized as sprite, stage, variable, list, clone, or event paths. Numeric comparisons use tolerances of $10^{-3}$ for sprite kinematics and $10^{-6}$ for other numeric fields; the looser kinematic tolerance absorbs floating-point jitter in continuous motion updates, while strings, costumes, Booleans, and list elements are matched exactly. Instances whose gold delta is empty contain no positive fields and are reported through no-op accuracy rather than through $F_1$. Outputs that fail schema parsing are scored as incorrect and logged as a distinct error category, so that formatting failure remains separable from simulation error in scorer reports.

\paragraph{Task 2: Causal event attribution.}
Given a Scratch world, an initial state $s_t$, an $H$-action sequence $a_{t:t+H-1}$, and a target event $q$, the model predicts whether the event occurs and identifies the responsible rule chain:
\[
(\hat{y}, \hat{c}) = f(\mathcal{W}, s_t, a_{t:t+H-1}, q),
\]
where $\hat{y}$ denotes event occurrence and $\hat{c}$ denotes the predicted causal chain. The ground truth $(y,c)$ is extracted from the runtime trace and executed block sequence. We evaluate event prediction using accuracy and $F_1$, causal attribution using exact rule-chain match against the trace-derived gold chain, and no-hallucination by measuring the rate at which a model abstains from inventing a causal chain when the event does not occur. Exact chain match is deliberately strict: it credits only a fully trace-faithful chain but does not localize where within a chain a prediction diverges, and graded chain credit (prefix or edit-distance scoring) is a natural refinement of this metric. The task separates event detection from causal attribution.

\paragraph{Task 3: Counterfactual prediction.}
Given an observed trajectory and an intervention $\iota$, the model predicts the outcome under the modified execution:
\[
\hat{s}^{\iota}_{t+H} = f(\mathcal{W}, s_t, a_{t:t+H-1}, \iota).
\]
The prediction is compared against the counterfactual ground truth $s^{\iota}_{t+H}$ obtained by replaying the intervened execution. Interventions may replace an action, move a sprite, change a threshold, modify a variable value, or alter an event condition. We report value-aware and presence-only changed-field $F_1$ on counterfactual outcomes with non-empty deltas, and no-op accuracy when replay shows that the intervention leaves all listed fields unchanged. Reporting the two pools separately prevents a system that always predicts change, or always predicts invariance, from appearing uniformly strong. The task measures consequence prediction under controlled changes.

\paragraph{Task 4: Long-horizon state tracking.}
Given an initial state $s_t$ and an action sequence $(a_t,\ldots,a_{t+H-1})$, the model predicts the final state without intermediate ground truth:
\[
\hat{s}_{t+H} = f(\mathcal{W}, s_t, a_{t:t+H-1}).
\]
We evaluate horizons such as $H \in \{2,5,10,15,20\}$. Metrics include full-state field accuracy, hidden-state accuracy, exact full-state match, and degradation as $H$ increases. The task exposes temporal drift when hidden variables, delayed broadcasts, clone states, or repeated collisions affect later behavior.

\paragraph{Input modalities and claim scope.}
Input modalities define how the executable world is presented. \sw{} supports five: raw project JSON, structured state tables, natural-language summaries, rendered screenshots, and short execution clips. The experiments below report primarily on the structured-state modality, under which the current state and action are given while the program rules are hidden. We adopt this partial-observation condition as the primary comparison because it is the regime in which the copy confound is sharpest, and because it measures a capability that matters when no executable specification is available, acting from observed state and action history alone. Under raw-program evidence, next-state prediction could in principle be solved by literally interpreting the program text, so we treat that condition as an upper-bound companion (see Discussion). Claims are therefore stated relative to the evidence condition, and the matched-sheet state-only versus state$+$program ablation in the Experiments section disaggregates missing information from simulation error on the same underlying sheet.

\paragraph{Splits and leakage control.}
We partition the benchmark at the project level into development and test splits. All instances derived from the same Scratch project remain in the same split, preventing project-level leakage across development and test.

\section{Experiments}
\label{sec:experiments}

We organize the evaluation around four research questions, one per task: \textbf{RQ1} asks whether a model can compute the executed one-step consequence of an action; \textbf{RQ2} asks whether it can recover the executed rule chain behind an event; \textbf{RQ3} asks whether it can compute the consequence of an intervention; and \textbf{RQ4} asks whether state tracking survives long horizons without intermediate ground truth. An ablation study attached to RQ1 asks why one-step prediction fails, varying evidence (state-only versus state$+$program), reasoning effort, and agent-mode tool use on the same synthetic sheet; RQ2--RQ4 are evaluated on their dedicated diagnostic slices.

\para{Models and data.}
We evaluate seven primary models, with coverage exceptions stated below: \textsc{Opus-5}, \textsc{GPT-5.6-Sol}, \textsc{DeepSeek-V4-Flash}, \textsc{DeepSeek-V4-Pro}, \textsc{MiMo-V2.5-Pro}, \textsc{Qwen3.7-Plus}, and \textsc{Kimi-K3}. For providers with an explicit reasoning-effort control, the main run uses the highest available reasoning setting. Local tools are disabled in the main comparison. Prompts never contain the target, gold label, or verification metadata; outputs that fail schema parsing are scored as incorrect.

The main synthetic sheet contains $160$ instances and the real-corpus sheet contains $252$ (Table~\ref{tab:dataset-map}). These are \emph{mixed task sheets}: changed-field $F_1$ is computed only on non-empty one-step deltas, no-op accuracy only on empty one-step deltas, and full-state metrics only on long-horizon items. We therefore report the applicable pool for each metric rather than treating every score as if it had denominator $160$. The diagnostic slices contain $120$ causal, $160$ counterfactual, and $200$ rollout instances.

\begin{table}[t]
\caption{Composition of the two mixed task sheets. The real-corpus sheet is larger than the synthetic sheet; task-specific metrics use only the applicable row. Seven models are reported, with task-specific coverage caveats stated for the affected rows.}
\label{tab:dataset-map}
\centering
\small
\setlength{\tabcolsep}{5pt}
\renewcommand{\arraystretch}{1.05}
\begin{tabular}{@{}lrr@{}}
\toprule
& Synthetic & Real \\
\midrule
Mechanism families / projects & 10 & 14 \\
One-step: changed delta & 17 & 72 \\
One-step: empty delta & 50 & 10 \\
Long-horizon & 93 & 170 \\
\midrule
Total & 160 & 252 \\
\bottomrule
\end{tabular}
\end{table}

\para{Metrics and reporting scope.}
The primary one-step metric is value-aware changed-field $F_1$ ($F_1^{\mathrm{VA}}$), with presence-only $F_1$ ($F_1^{\mathrm{pres}}$) separating field selection from value computation. Empty deltas are evaluated by no-op accuracy. Long-horizon predictions use full-state field accuracy, hidden-state accuracy, and exact full-state match. The synthetic changed-delta pool contains only $17$ instances, so small point differences should not be interpreted as statistically established rankings.

\para{RQ1: one-step computation (Task 1, next-state).}
Under state-only evidence, seven models answer the synthetic one-step items from the current structured state and action, with the program hidden. Table~\ref{tab:main-results} reports changed-delta $F_1$ and empty-delta accuracy. Value-aware $F_1$ spans $0.0$--$30.0\%$: \textsc{Opus-5} and \textsc{Kimi-K3} form the leading tier; \textsc{GPT-5.6-Sol} reaches $14.0\%$; and four systems remain at or below $5\%$. No-op accuracy varies independently of changed-delta skill: \textsc{Kimi-K3} pairs the second-best changed-delta $F_1$ with the worst no-op accuracy ($46\%$), over-predicting change on invariant cases.

The presence--value comparison reveals two different failure modes. For six of seven models, $F_1^{\mathrm{pres}}=F_1^{\mathrm{VA}}$: whenever these systems name a correct changed field, they also compute its value, so their entire loss comes from field selection. The strongest row, \textsc{Opus-5}, additionally shows value-computation error after field selection ($40.0$ presence-only versus $30.0$ value-aware). The bottleneck is therefore tier-dependent rather than a single universal ``computation gap.''

\finding{RQ1}{State-only prediction fails at field selection for most models, and at value computation even for the strongest.}

\begin{table}[t]
\caption{State-only one-step prediction on the synthetic sheet (\%). $F_1$ uses the $17$ changed-delta items and no-op accuracy the $50$ empty-delta items; all rows use the complete task-specific pools.}
\label{tab:main-results}
\centering
\small
\setlength{\tabcolsep}{4.5pt}
\renewcommand{\arraystretch}{1.02}
\begin{tabular}{@{}lrrr@{}}
\toprule
Model & VA $F_1$ & Pres. $F_1$ & No-op \\
\midrule
\textsc{Opus-5}             & \textbf{30.0} & 40.0 & 94 \\
\textsc{Kimi-K3}            & 27.3 & 27.3 & 46 \\
\textsc{GPT-5.6-Sol}        & 14.0 & 14.0 & 80 \\
\textsc{DeepSeek-V4-Pro}    & 5.0  & 5.0  & 84 \\
\textsc{MiMo-V2.5-Pro}      & 4.9  & 4.9  & 60 \\
\textsc{DeepSeek-V4-Flash}  & 0.0  & 0.0  & 100 \\
\textsc{Qwen3.7-Plus}       & 0.0  & 0.0  & 94 \\
\bottomrule
\end{tabular}
\end{table}

\para{RQ1 anchors: what trivial strategies achieve.}
The no-change prior and deterministic action heuristic establish two different shortcuts (Table~\ref{tab:metric-diagnostic}). Copying receives no changed-field credit, but retains $91.4\%$ field accuracy on the synthetic full-state pool because most fields persist. The apply-action heuristic reaches $10.3\%$ on controlled synthetic changed-delta items but $0\%$ on real projects. The VM-replay oracle reaches $100\%$ wherever the corresponding metric is defined. Because these anchors are scored on task-specific pools, they diagnose the metrics rather than set a bar that every prompted model must clear.

\begin{table}[t]
\caption{Model-free anchors (\%). Each anchor is scored only where it applies: VA $F_1$ on the non-empty changed-delta pools ($17$ synthetic, $72$ real) and full-state field accuracy (Full) on the $93$ synthetic long-horizon items. The action heuristic predicts one-step deltas rather than complete states, so Full does not apply (``--''); the copy prior scores $91.4\%$ Full despite $0\%$ VA $F_1$ because most fields persist on the synthetic long-horizon pool.}
\label{tab:metric-diagnostic}
\centering
\small
\setlength{\tabcolsep}{4pt}
\begin{tabular}{@{}llrr@{}}
\toprule
Method & Applicable pool & VA $F_1$ & Full \\
\midrule
Copy prior & synthetic + real & 0.0 & 91.4 \\
Action heuristic & synthetic  & 10.3 & -- \\
Action heuristic & real  & 0.0 & -- \\
VM replay & synthetic + real & 100.0 & 100.0 \\
\bottomrule
\end{tabular}
\end{table}

\para{RQ1 split: real versus synthetic.}
The real-corpus changed-delta pool is qualitatively harder. Six models cover all $72$ changed outcomes, while \textsc{Kimi-K3} covers $71$ because one oversized project exceeds its request limit. Every reported row nevertheless scores $0.0\%$ value-aware $F_1$. \textsc{Opus-5} still obtains $4.4\%$ presence $F_1$, and \textsc{GPT-5.6-Sol} obtains $0.9\%$: they occasionally identify a changed field but never compute its executed value. The other five systems receive no field credit. This result is stronger than a simple value error: most systems fail before value computation begins.

\finding{RQ1 split}{Controlled mechanism families admit partial rule inference, but real projects defeat value computation entirely and field selection almost entirely.}

\para{RQ1 ablation study.}
The remaining question for RQ1 is \emph{why} state-only prediction fails. Three controlled ablations on the same synthetic sheet each isolate one candidate explanation: missing rule information (evidence), insufficient test-time compute (reasoning effort), and missing procedural scaffolding (tool use).

\begin{figure}[t]
\centering
\includegraphics[width=\linewidth]{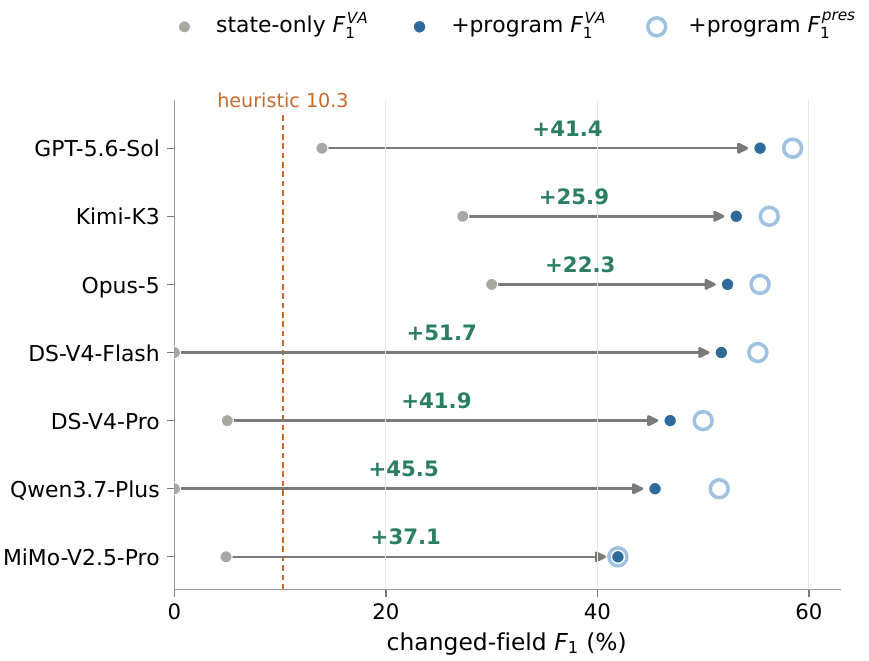}
\caption{Matched-sheet evidence ablation on the synthetic changed-delta pool. Arrows connect each model's state-only value-aware $F_1$ (gray) to its state$+$program score (blue), with the paired gain annotated; open rings mark presence-only $F_1$ under program evidence, which exceeds the value-aware score by at most $6.1$ points and coincides with it for \textsc{MiMo-V2.5-Pro}. All seven reported models cover all $17$ changed items in both conditions, so every comparison is strictly paired; the three \textsc{Qwen3.7-Plus} refusals lie outside the changed-delta pool. The deterministic heuristic is shown at $10.3\%$; the VM oracle at $100\%$ is off scale.}
\label{fig:evidence-ladder}
\end{figure}

\para{Ablation 1 (evidence): information deficit or execution failure?}
The matched-sheet evidence ablation repeats the synthetic sheet with the full program view added to the prompt. Program access raises every reported model's value-aware score by $22.3$--$51.7$ percentage points (Figure~\ref{fig:evidence-ladder}), confirming a substantial information deficit in the state-only condition. Across the reported rows, it also compresses the cross-model range from $30.0$ points under state-only evidence ($0.0$--$30.0\%$) to $13.4$ points using unrounded scores with the program visible ($41.9$--$55.4\%$), while changing the ordering: \textsc{GPT-5.6-Sol} moves from third to first, and \textsc{DeepSeek-V4-Flash} moves from the tied last tier to fourth. State inference and explicit program following are therefore distinct capabilities that rank the models differently.

Under state$+$program evidence, value-aware and presence-only scores track closely: their difference is $0$--$6.1$ points for every model (filled versus open markers in Figure~\ref{fig:evidence-ladder}). Value errors after correct field selection therefore explain little of the remaining deficit. The important residual is instead an \emph{execution gap}: even with the rule visible, the best model reaches only $55.4\%$ against the $100\%$ VM oracle, so errors still include missing or extra fields as well as incorrect values.

\para{Ablation 2 (reasoning effort): is more test-time compute monotonic?}
Holding the prompt and evidence fixed, we sweep the provider reasoning control for \textsc{Opus-5} and \textsc{GPT-5.6-Sol} (Table~\ref{tab:reasoning-budget}). Value-aware $F_1$ is non-monotonic for both: \textsc{Opus-5} dips at \texttt{medium} before recovering, while \textsc{GPT-5.6-Sol} peaks at \texttt{high} ($20.0\%$) and falls at \texttt{xhigh} ($14.0\%$). No-op accuracy should not be read as ``caution'' or answer rate; it is correctness on the available invariant cases. This task-specific non-monotonicity is consistent with recent reports that additional test-time reasoning need not improve accuracy monotonically~\citep{ghosal2025thinking}, although our two-model sweep does not test the mechanism proposed there. The no-op trajectories show that effort also changes the balance between predicting changes and preserving invariance: \textsc{Opus-5} rises from $56\%$ to $94\%$, whereas \textsc{GPT-5.6-Sol} falls from $100\%$ to $80\%$ and then plateaus.\looseness=1

\begin{table}[t]
\caption{Reasoning-effort sweep (\%). Columns \mbox{1--5} are low\slash medium\slash high\slash xhigh\slash max for \textsc{Opus-5}, and none\slash low\slash medium\slash high\slash xhigh for \textsc{GPT-5.6-Sol}. Some \textsc{Opus-5} intermediate runs contain $155$ rather than $160$ items because of provider refusals; their no-op denominators are therefore $48$--$49$ rather than $50$.}
\label{tab:reasoning-budget}
\centering
\small
\setlength{\tabcolsep}{3.0pt}
\begin{tabular}{@{}llrrrrr@{}}
\toprule
Model & Metric & 1 & 2 & 3 & 4 & 5 \\
\midrule
\multirow{2}{*}{\textsc{Opus-5}} & VA $F_1$ & 24.0 & 20.3 & 25.9 & 29.1 & 30.0 \\
& No-op & 56 & 69 & 80 & 86 & 94 \\
\addlinespace[2pt]
\multirow{2}{*}{\textsc{GPT-5.6-Sol}} & VA $F_1$ & 0.0 & 0.0 & 10.0 & 20.0 & 14.0 \\
& No-op & 100 & 88 & 82 & 80 & 80 \\
\bottomrule
\end{tabular}
\end{table}

\para{Ablation 3 (tool use): do generic tools substitute for missing dynamics?}
The agent condition supplies shell and file tools but does \emph{not} reveal the hidden Scratch program or provide the pinned Scratch VM\@. We therefore use it as a generic tool-use skyline rather than an execution skyline: it tests whether procedural tools help organize the available evidence, not whether the system can replay the world. The effect is model-dependent (Table~\ref{tab:tool-ablation}). For \textsc{GPT-5.6-Sol}, agent mode lowers value-aware $F_1$ from $14.0$ to $10.5\%$: precision rises ($30.0\to40.0\%$), but recall falls ($9.1\to6.1\%$), so the agent becomes more selective and misses more true changes. For \textsc{Opus-5}, agent mode raises $F_1$ from $30.0$ to $36.9\%$ through gains in both precision and recall, but no-op accuracy falls from $94$ to $86\%$. Generic tools thus shift the error profile in opposite directions rather than providing a uniform capability gain.

\begin{table}[t]
\caption{Generic tool-use ablation on the synthetic one-step pools (\%). Agent mode supplies shell and file tools but not the hidden Scratch program or pinned Scratch VM.}
\label{tab:tool-ablation}
\centering
\small
\setlength{\tabcolsep}{3.0pt}
\begin{tabular}{@{}llrrrr@{}}
\toprule
Model & Mode & VA $F_1$ & Prec. & Recall & No-op \\
\midrule
\multirow{2}{*}{\textsc{GPT-5.6-Sol}} & tool-free & 14.0 & 30.0 & 9.1 & 80 \\
& agent & 10.5 & 40.0 & 6.1 & 84 \\
\addlinespace[2pt]
\multirow{2}{*}{\textsc{Opus-5}} & tool-free & 30.0 & 33.3 & 27.3 & 94 \\
& agent & 36.9 & 37.5 & 36.4 & 86 \\
\bottomrule
\end{tabular}
\end{table}

\finding{ablation study}{One-step failure reflects both missing information and execution failure: program access reduces the deficit substantially yet leaves a $44.6$-point execution gap, and neither higher reasoning effort (non-monotonic) nor generic tools (opposite, model-specific shifts) close it.}

\begin{figure}[th!]
\centering
\includegraphics[width=\linewidth]{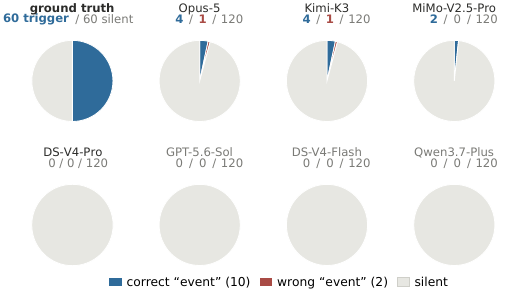}
\caption{Causal event detection fails silently. Each pie splits a model's $120$ causal cases by behavior: blue marks predictions of ``event'' where a real event occurred (correct), red marks predictions of ``event'' where none occurred (wrong), and gray marks silence, where the model predicts no event; labels read correct$\,/\,$wrong$\,/\,$total. The ground-truth pie shows what a perfect detector would do: trigger on the $60$ event cases (blue half) and stay silent on the other $60$ (gray half). The models instead trigger on at most $5$ of $120$ cases (a $0$--$4.2\%$ trigger rate against the $50\%$ base rate), four stay entirely silent, and no exact executed chain is ever recovered.}
\label{fig:causal}
\end{figure}

\para{RQ2: mechanism recovery (Task 2, causal attribution).}
The causal slice is balanced: $60$ event-positive and $60$ event-negative instances, so a constant no-event predictor already scores $50\%$ accuracy. The evaluated models fail \emph{silently} (Figure~\ref{fig:causal}). Across their $840$ predictions, the seven models say ``event'' only $12$ times, $10$ of them correctly: \textsc{Opus-5} and \textsc{Kimi-K3} detect $4$ of the $60$ real events, \textsc{MiMo-V2.5-Pro} detects $2$, and the remaining four detect none; four never trigger at all. False triggers are just as rare ($\leq 1$ per model; \textsc{MiMo-V2.5-Pro} additionally invents $2$ causal chains without flagging an event). The superficially high no-hallucination rates ($96.7$--$100\%$) are therefore a by-product of silence rather than reliability, and event accuracy ($50.0$--$52.5\%$) remains within 2.5 points of the baseline. Across the three models that ever predict an event, $10$ of their $12$ positive event predictions are correct, indicating that severe under-triggering, rather than excessive false positives, is the dominant failure mode; nevertheless, no model recovers an exact executed rule chain (exact-chain match is $0$ wherever defined and undefined for models that never predict an event).\looseness=1

\finding{RQ2}{Causal attribution fails silently: models trigger on $1.4\%$ of cases in a slice where half contain a real event, and never recover an exact executed chain.}

\begin{figure}[t]
\centering
\includegraphics[width=\linewidth]{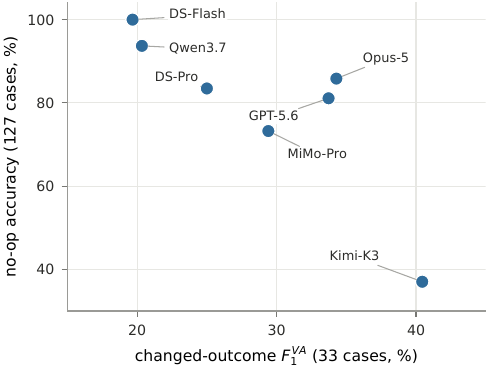}
\caption{Counterfactual selectivity. Changed-outcome $F_1^{\mathrm{VA}}$ ($33$ cases) against no-op accuracy ($127$ no-op cases), one point per model. The empty upper-right corner and the negative association (Spearman $\rho=-0.75$) show that no model is strong on both axes.}
\label{fig:cf}
\end{figure}

\para{RQ3: intervention consequences (Task 3, counterfactual).}
The $160$ counterfactual instances contain $33$ changed outcomes and $127$ no-op outcomes. Figure~\ref{fig:cf} exposes a selectivity trade-off that a single aggregate score would hide. \textsc{Kimi-K3} has the highest changed-outcome $F_1^{\mathrm{VA}}$ ($40.4\%$) but only $37.0\%$ no-op accuracy. \textsc{DeepSeek-V4-Flash} is the opposite extreme: $19.7\%$ changed-outcome $F_1$ and $100\%$ no-op accuracy. \textsc{Opus-5} and \textsc{GPT-5.6-Sol} are more balanced ($34.3/85.8$ and $33.7/81.1$, respectively). Across the seven models, the two metrics have a descriptive Spearman association of $\rho=-0.75$; no model dominates both consequence computation and no-op recognition.

\begin{figure*}[!t]
\centering
\includegraphics[width=\textwidth]{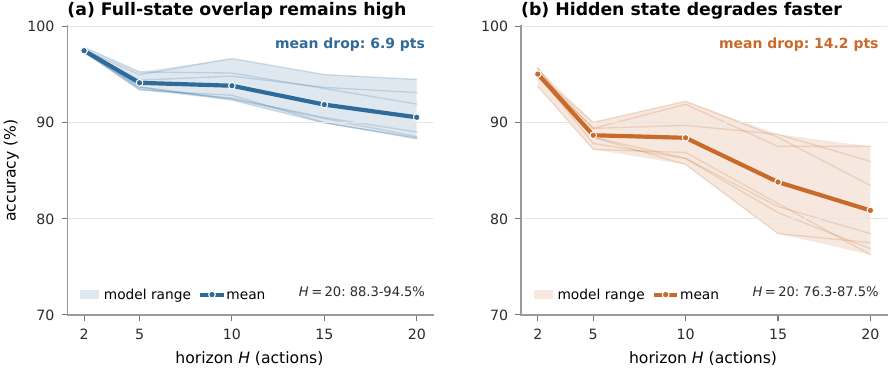}
\caption{Long-horizon state tracking across seven reported models. The thick line is the cross-model mean and the ribbon spans the model range. (a) Mean full-state field accuracy drops $6.9$ points from $H{=}2$ to $H{=}20$. (b) Mean hidden-state accuracy drops $14.2$ points, approximately twice as much.}
\label{fig:rollout}
\end{figure*}

\finding{RQ3}{Counterfactual errors are asymmetric: models either over-predict change or over-preserve invariance, so changed and no-op outcomes must be reported separately.}

\para{RQ4: long-horizon tracking (Task 4, rollout).}
On the dedicated $200$-instance rollout slice, seven models predict final states at $H\in\{2,5,10,15,20\}$ without intermediate ground truth. Across models, mean full-state accuracy drops by $6.9$ points from $H{=}2$ to $H{=}20$, while mean hidden-state accuracy drops by $14.2$ points, about twice as much (Figure~\ref{fig:rollout}). At $H{=}20$, full-state accuracy remains $88.3$--$94.5\%$, but hidden-state accuracy falls to $76.3$--$87.5\%$; aggregate exact full-state match ranges from $38.0$ to $62.0\%$ across models.

\finding{RQ4}{Temporal error concentrates in hidden state, which degrades roughly twice as fast as full-state overlap.}

\section{Discussion}
\label{sec:discussion}

\paragraph{Metric lesson.}
Executable prediction should be decomposed into three questions: does the state largely persist, are the affected fields identified, and are their executed values correct? Full-state field accuracy, $F_1^{\mathrm{pres}}$, and $F_1^{\mathrm{VA}}$ answer these three questions in turn. The experiments further show that the location of the bottleneck changes with model and evidence. Under state-only evidence, lower-tier systems fail field selection while the leading systems show an additional value error. Under state$+$program evidence, presence and value scores differ by at most $6.1$ points, so the remaining distance to the oracle is a broader execution failure rather than a universal value-computation gap.

\paragraph{Evidence changes what is being ranked.}
Program access does more than raise scores: across the reported rows, it compresses the cross-model range from $30.0$ to $13.4$ points using unrounded scores and substantially reorders models. A system that is strong at inferring hidden dynamics from state can be weaker at following an explicit block program, and vice versa. Consequently, ``best world model'' is underspecified unless the evidence condition is fixed. State-only evaluation probes prediction with incomplete rules; state$+$program evaluation probes program-conditioned execution; the VM oracle provides the exact executable upper bound.\looseness=1

\paragraph{Selectivity is a first-class error axis.}
The causal and counterfactual slices show why accuracy or a single aggregate $F_1$ can be misleading. On the balanced causal set, near-$50\%$ accuracy is explained by the no-event baseline rather than successful event detection. In counterfactual prediction, the model with the strongest changed-outcome score performs worst on invariant outcomes, while another model shows the reverse. Generic tool use produces the same kind of shift: \textsc{GPT-5.6-Sol} gains precision but loses recall, whereas \textsc{Opus-5} gains changed-field recall at the cost of no-op accuracy. Reporting positive and invariant cases separately reveals these trade-offs.

\paragraph{Benchmark use.}
\sw is most informative when systems are compared by error profile rather than by a single leaderboard number. A program parser may improve field selection and causal traces; a learned transition model may improve value updates on repeated mechanisms; a visual model may improve scene recovery while still missing hidden state. Shared replay sources and canonical schemas make these differences comparable across rule access, state update, intervention handling, and temporal persistence.

\paragraph{Relation to tools and execution.}
An agent with the pinned Scratch VM and complete project can answer \sw queries by execution; this system defines the $100\%$ VM oracle. The reported models instead operate without direct access to the executor, so their scores measure reasoning over evidence rather than literal simulation. The tool-use ablation therefore bounds the effect of generic tools under the same state-only evidence; it does not evaluate access to an executor. The opposite effects for \textsc{Opus-5} and \textsc{GPT-5.6-Sol} show that generic tools can alter search and selectivity without recovering missing dynamics. Under state$+$program evidence, a literal interpreter is a conceptual upper bound, but prompted models stop at $55.4\%$ and do not behave like exact interpreters.

\section{Limitations}
\label{sec:limitations}

\sw targets action-conditioned executable state prediction, not general physical understanding, embodied control, or open-ended video prediction. Its worlds are program-defined: hidden variables, lists, broadcasts, clones, and event handlers are executable state variables rather than latent physical quantities. This precision enables replay verification but limits the scope of the conclusions.

The principal statistical limitation is the size of several task-specific pools. The synthetic changed-delta result uses $17$ instances, even though the mixed sheet contains $160$ total items. Causal results use $60$ positive events, and the tool comparison covers only two reported models. We therefore treat one- or two-point differences as descriptive and emphasize large tier, split, and error-profile patterns. Provider refusals and context limits also produce minor coverage differences: \textsc{Kimi-K3} completes $250/252$ real items, while ten \textsc{MiMo-V2.5-Pro} long-horizon timeouts are conservatively scored as failures.

The current analysis computes changed-field and full-state metrics on different task-specific pools. Thus, the $0\%$ copy score and $91.4\%$ full-state overlap are a cross-pool metric diagnostic rather than a same-instance estimate. Future work should report same-instance copy overlap for every one-step instance and project-macro confidence intervals.

Causal labels require special care because Scratch concurrency, clones, and nested broadcasts can make ordered explanations ambiguous. We exclude instances whose chains cannot be recovered reliably, and exact chain match is deliberately strict. Graded prefix or edit-distance credit would localize errors that the current metric only bounds. The present models also have near-zero positive event recall, so stronger event detectors are needed before fine-grained mechanism attribution can be meaningfully compared.

The evidence ablation uses synthetic projects whose programs fit in the prompt; extending this analysis to larger real projects may require retrieval or structured program representations. Our conclusions therefore concern zero-shot prompted models, while whether trained transition models can close the execution gap remains open. Human reference performance, trained dynamics baselines, same-project macro scores, and broader real-project coverage remain important extensions. Treating unparseable outputs as incorrect is conservative; the score blocks retain parse-failure counts, so formatting errors remain separable from simulation errors.\looseness=1

\section{Conclusion}
\label{sec:conclusion}

We introduced \sw, an execution-grounded diagnostic benchmark for action-conditioned world models in executable Scratch worlds. Replay-verified labels and value-aware changed-field scoring reveal when high state overlap reflects true execution rather than persistence alone. The experiments expose a structured execution gap: state-only prediction achieves strong overlap but weak executed-change accuracy, and even with full program access the strongest model reaches only $55.4\%$ against the $100\%$ VM oracle. On real projects, no reported model computes a correct changed value, showing that plausible state prediction does not imply rule execution.
These diagnostics position \sw not as a single leaderboard, but as a framework for identifying whether a system lacks rules, misses changed fields, computes incorrect values, fails to capture causal structure, or loses hidden state over time. Because every label is a replay outcome, the same pipeline extends to new mechanisms, larger corpora, and trained predictors. Closing this execution gap is therefore a concrete milestone toward world models that genuinely compute consequences rather than merely predict plausible states.

\clearpage
\balance
\bibliographystyle{ACM-Reference-Format}
\bibliography{refs}

\end{document}